\documentclass[12pt]{iopart}
\usepackage{iopams}
\usepackage{mathrsfs}
\usepackage{amssymb}
\usepackage{graphicx,epstopdf,epsfig}
\usepackage{bm}
\usepackage{color}
\usepackage{times}
\usepackage{hyperref}

\begin{document}

\title{Demonstration of an inductively coupled ring trap for cold atoms}

\author{J. D. Pritchard, A. N. Dinkelaker, A. S. Arnold, P. F. Griffin and E. Riis}
\ead{e.riis@strath.ac.uk}
\address{Department of Physics, SUPA, University of Strathclyde, Glasgow G4 0NG, United Kingdom}

\date{\today}

\begin{abstract}
We report the first demonstration of an inductively coupled magnetic ring trap for cold atoms. A uniform, ac magnetic field is used to induce current in a copper ring, which creates an opposing magnetic field that is time-averaged to produce a smooth cylindrically symmetric ring trap of radius 5~mm. We use a laser-cooled atomic sample to characterise the loading efficiency and adiabaticity of the magnetic potential, achieving a vacuum-limited lifetime in the trap. This technique is suitable for creating scalable toroidal waveguides for applications in matterwave interferometry, offering long interaction times and large enclosed areas.
\end{abstract}

\maketitle

\section{Introduction} 

Development of ring traps for cold atoms is an active topic of theoretical and experimental study, motivated by the ability to create one-dimensional waveguides with periodic boundary conditions, which have applications in two regimes. For radii smaller than $\sim100~\mu$m these traps can be filled with quantum degenerate gases, enabling studies of persistent current flow of a superfluid in a multiply connected geometry to realise atomtronic \cite{seaman07} analogues to a SQUID \cite{ramanathan11,moulder11} or of Hawking radiation from sonic black holes \cite{garay00}. Alternatively, large radius rings can be utilised to create a matter-wave interferometer \cite{cronin09}. Atom interferometry is sensitive to both inertial effects and external fields and has been used to perform precision measurements of rotation \cite{gustavson97,lenef97,wu07},  acceleration and gravitation \cite{canuel06,su10}, in addition to determination of fundamental constants \cite{gupta02}, magnetic gradients \cite{wang05} and ac Stark shifts \cite{deissler08}. State-of-the-art experiments typically use unguided atoms, requiring large path lengths at which point acceleration due to gravity or the coriolis force due to the earths rotation becomes significant \cite{lan12}. Ring traps provide an ideal geometry for these applications due to the common-mode rejection of the identical paths, with trivial extension of the enclosed area using multiple revolutions \cite{wu07}. They are especially suited to performing rotation measurements using the Sagnac effect \cite{sagnac13}. As the rotation sensitivity of a Sagnac interferometer is directly proportional to the enclosed area, large area interferometers are desirable \cite{arnold06}. 

Early techniques for generating large area ring waveguides exploited dc magnetic traps to create large ring \cite{arnold06,sauer01} and stadium \cite{wu04} geometries. One of the challenges in producing scalable magnetic traps is avoiding losses from Majorana spin-flips at field zeros, which can be achieved using time-averaged magnetic fields \cite{arnold04,gupta05} or rf dressing \cite{zobay01,lesanovsky06,vangeleyn12}. These methods can be combined to realise time-averaged adiabatic potentials (TAAP) \cite{lesanovsky07,sherlock11} to create versatile traps with adaptable radii. For studies of persistent flow in quantum fluids small traps with $r\sim20~\mu$m are required which are possible using combined magnetic and optical or all optical traps \cite{moulder11,ryu07,bruce11,henderson09}. A novel technique to create time-averaged toroidal potentials using mechanical oscillation of a magnetic nanowire has recently been proposed \cite{west12}.

In this paper we demonstrate a large radius ring trap created from the time-averaged potential arising from the current induced in a conducting ring \cite{griffin08}, suitable for applications in the regime of atom interferometry. A key benefit to this trapping scheme is that the geometry is defined by a macroscopic circular conductor, avoiding end effects associated with dc electromagnetic traps that act to break the cylindrical symmetry of the ring \cite{crookston05}. The other advantage is that the ac field averages out any roughness in the magnetic field to create a smooth trapping potential.

\section{Theory}\label{sec:theory}

\begin{figure}
\flushright
	\includegraphics{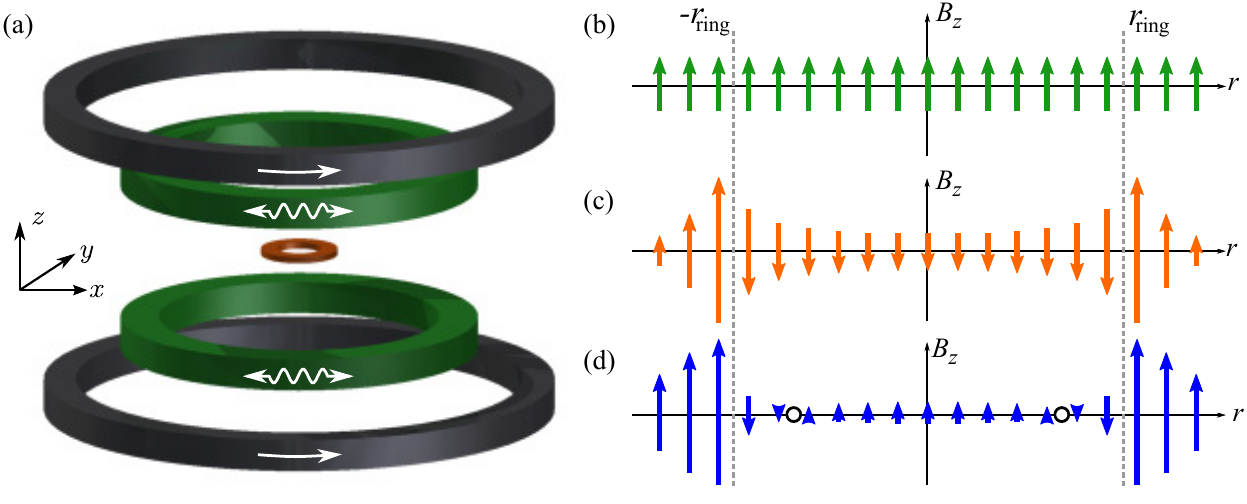}
	\caption{Inductive Ring Trap. (a) A small copper ring is placed at the centre of two pairs of Helmholtz coils that provide an ac magnetic field at frequency $\omega$ (green coils) and a static bias field (grey coils) along the $\hat{z}$-axis. Gravity acts along $-\hat{z}$. (b)-(d) show the magnetic field along $\hat{z}$ in the plane of the copper ring at time $t=0$, where (b) is the ac drive field and (c) is the magnetic field created by the out of phase induced current in the ring. The total field is shown in (d), resulting in a field zero indicated by $\circ$ that is time-averaged to give a magnetic ring trap.}
	\label{fig:fig1}
\end{figure}

An inductively coupled ring trap for cold atoms is realised using the configuration shown schematically in \fref{fig:fig1}(a), where a small conducting ring of radius $r_\mathrm{ring}$ is placed at the centre of two pairs of Helmholtz coils, one of which produces a time-varying ac drive field $\bm{B}_1(t) = B_0\cos(\omega t)\bm{\hat{z}}$ and another providing a uniform dc bias field $\bm{B}_2 = B_\mathrm{b}\bm{\hat{z}}$ both perpendicular to the plane of the ring. The time-varying magnetic field induces a current in the the ring proportional to the rate of change of the magnetic flux through the ring, which following Griffin \textit{et al.}~\cite{griffin08} is given by
\begin{equation} \label{eq:I}
I(t) = -\frac{\pi r^2_\mathrm{ring}B_0}{L\sqrt{1+\Omega^{-2}}}\cos(\omega t + \delta_0),
\end{equation}
where $R$ and $L$ are the electrical resistance and inductance of the ring, $\Omega = \omega L/R$ is the ratio of the ac drive frequency to the natural low-pass frequency of the ring and $\delta_0 = \tan^{-1}(1/\Omega)$ is the phase-shift of the induced current. The current induced in the ring creates a spatially inhomogeneous magnetic field $\bm{B}_\mathrm{ring}(r,z)$, as illustrated schematically in \fref{fig:fig1}(c). Inside the ring this field opposes the drive field $\bm{B}_1(t)$ in \fref{fig:fig1}(b) such that at any time in the cycle the total instantaneous magnetic field vanishes on a circle inside the ring radius, as indicated in \fref{fig:fig1}(d).

\begin{figure}
	\flushright
	\includegraphics{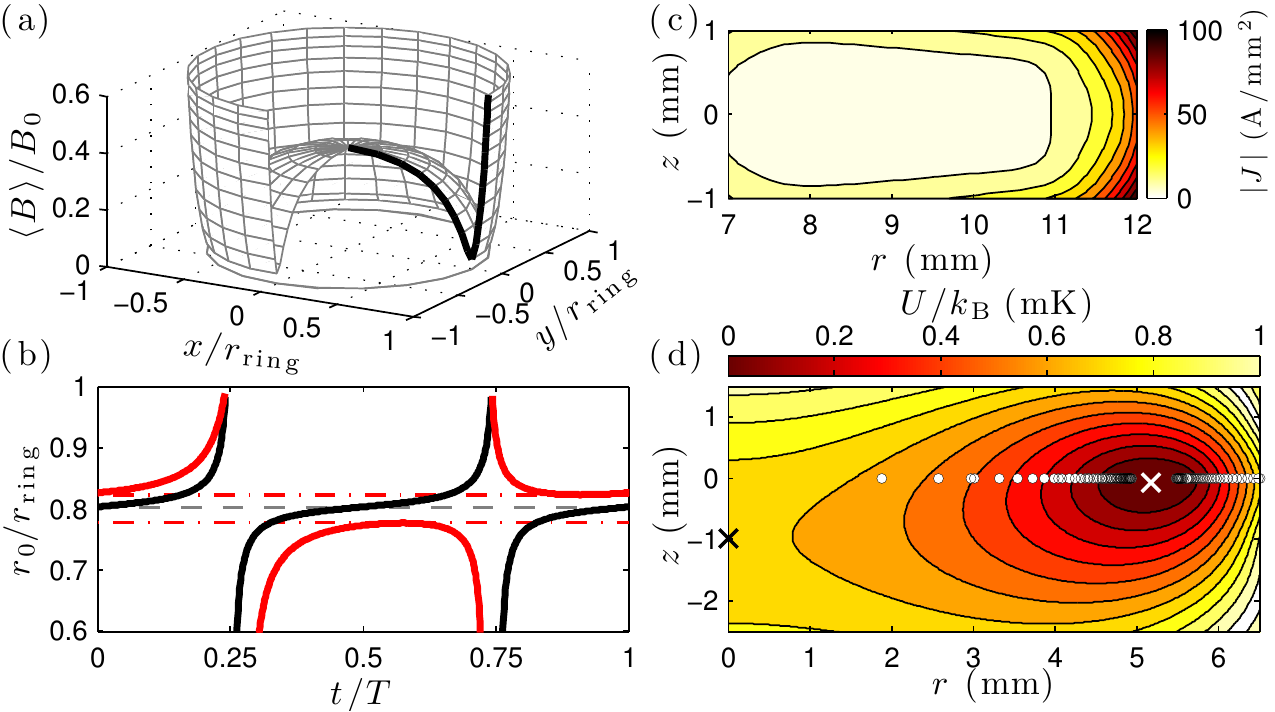}
	\caption{Time-Averaged Ring Potential (a) Cycle-averaged B-field magnitude in the plane of the ring creating an axially symmetric minimum inside the conductor radius. (b) Radius of the zeros in the instantaneous magnetic field during the ac cycle, calculated for $\Omega=20$ with no external bias (black) and for $B_\mathrm{b}=0.1B_0$ (red) to show the zeros being excluded from the trap centre at $r/r_\mathrm{ring}=0.8$ due to the bias field, preventing Majorana spin-flips. (c) Finite element simulation of the induced current density within the cross-section of the copper ring used in the experiment for $B_0=110$~G at $\omega/2\pi=30~$kHz, showing the current is strongly localised to the outer edge of the ring. (d) Time-averaged trapping potential for $\vert F=2, m_F=+2\rangle$ state of Rb including gravity, calculated from the current distribution in (c) for $B_\mathrm{b}=5$~G. White circles indicate positions of field-zeros across the cycle, white cross indicates trap minimum and the black cross marks the saddle point used to define the trap-depth.}
	\label{fig:fig2}
\end{figure}

If the ac drive frequency $\omega$ is much larger than the frequency of atomic motion within the trap (typically $\sim100$~Hz for magnetic traps), the effective magnetic field experienced by an atom is given by the field magnitude averaged over one cycle $\langle {B} \rangle = \frac{1}{T}\int_T\vert\bm{B}(t)\vert\mathrm{d}t$ \cite{petrich95,arnold04,griffin08}, where $T=2\pi/\omega$ is the cycle period. The resulting time-averaged field magnitude creates an cylindrically symmetric trapping potential $U = m_Fg_F \mu_B \langle {B} \rangle$ for atoms in weak-field-seeking states with $m_Fg_F>0$ that provides both radial and axial confinement to create a toroidal trap. This is seen from \fref{fig:fig2}(a) which plots $\langle {B} \rangle$ in the plane of the ring ($z=0$) calculated for $\Omega=20$, with a steep barrier on the outside of the trap due to the large magnetic field close to the conducting ring and a parabolic local maximum at the origin. Minimisation of $\langle B \rangle$ with respect to $r$ shows the radius of time-averaged minimum $r_\mathrm{trap}$ is located at the point where the inhomogeneous field of the ring is equal to $\bm{B}_\mathrm{ring}(r_\mathrm{trap},0)=-B_0\cos\delta_0\bm{\hat{z}}$. As the induced field $\bm{B}_\mathrm{ring}\propto B_0$, the trap radius is determined purely by the conductor geometry. In addition, for $\Omega\gg1$ the induced current is independent of resistance. Thus the trap radius is robust against fluctuations in the external field amplitude and the effects of Ohmic heating. The ac current also averages out corrugations caused by current meandering within the conductor \cite{kraft02,leanhardt03} to create a smooth, symmetric potential.

In order for the atomic magnetic dipole to follow the external field adiabatically it is necessary for the Larmor frequency $\omega_L \propto \mu_B B$ to be much larger than the rate of change of the magnetic field given by $\dot{\theta} = \vert \dot{\bm{B}}\times\bm{B}\vert/\vert \bm{B}\vert^2$ \cite{rabi54}. If this condition is not met, then the atom can undergo a Majorana spin-flip into an un-trapped magnetic spin state and is lost from the ring. For the time-averaged potential it is the instantaneous value of the field, which must meet the adiabaticity criterion at all times in the cycle requiring $\vert \bm{B}(t)\vert>0$. \Fref{fig:fig2}(b) plots the radial coordinate of the instantaneous zero $r_0$ throughout the cycle, which shows that the magnetic field zeros spend the majority of the time centred at the trap minimum $r_\mathrm{trap}$, and sweep through the whole ring plane at times $t=T/4$ and $3T/4$ due to the phase-lag between the driving and induced fields. Thus atoms loaded into the trap would be rapidly lost due to the non-adiabatic potential within a few cycles of the ac field. It is therefore necessary to apply an external dc bias field to move the field-zeros away from the trap minimum. The trivial solution is to place a current carrying wire aligned along the $\hat{z}$ axis through the centre of the ring to generate a radially symmetric azimuthal magnetic field \cite{arnold06}. However, this approach has a number of drawbacks as it compromises optical access within the ring trap and the end effects of the wire can break the symmetry. It also reduces the scalability of the ring trap for creating traps with radii of a few mm.

Instead we consider the case of a uniform dc bias field $B_\mathrm{b}\hat{z}$ that is generated using a second pair of Helmholtz coils as shown in \fref{fig:fig1}(a), which acts to offset the ac field to remove the field zeros. A lower limit for the bias required to remove the zeros from the trap centre can be obtained analytically from the amplitude of the combined ac field at $r_\mathrm{trap}$ using the relation for $\bm{B}_\mathrm{ring}(r_\mathrm{trap},0)$ given above such that $B_\mathrm{b}>\vert B_0\left[\cos(\omega t)-\cos\delta_0\cos(\omega t+\delta_0)\right]\vert$, which simplifies to
\begin{equation}\label{eq:Bmin}
B_\mathrm{b}>\frac{B_0}{\sqrt{1+\Omega^2}},
\end{equation}
corresponding to 5.5 G for a 110 G drive field and $\Omega=20$. As the bias is increased, the circle of zero field is pushed out from the trap centre, creating inner and outer radii in the plane of the ring outside of which atoms are lost, analogous to the `circle of death' in a TOP trap \cite{petrich95}. The effect of the bias field can be seen clearly from \fref{fig:fig2}(b), where the zeros no longer sweep through the trap centre but create an exclusion region around the trap in which atoms can adiabatically follow the field. The relevant characteristic for the trap is therefore the equilibrium temperature which corresponds to the lowest potential energy at which the zeros occur during the cycle relative to the trap centre. A downside of the bias field is to offset the trap minimum, resulting in a reduction in trap depth and a weakening of the harmonic confinement in the radial direction. For large bias fields the trap becomes flat and anharmonic along $r$, requiring a compromise between tight trap frequencies and equilibrium temperatures greater than $\sim10\mu$K. For ultracold gases, such as a Bose-Einstein condensate, this is not an issue as only a weak bias field is required. An alternative solution proposed by Griffin \textit{et al.} \cite{griffin08} is to apply a quadrupole magnetic field centred on the ring, however for our present trapping parameters the gravitational sag due to the weak ($\sim100$~Hz) axial trap frequencies causes the resulting trap minimum to overlap with the shifted zeros.

Our experiment utilises a 2~mm thick copper ring with internal and external radii of 7 and 12~mm respectively, machined from an OHFC copper gasket which has been electropolished to give a smooth surface. These dimensions are chosen to provide a large thermal mass to prevent distortion of the ring due to heating from the induced current. However the large conductor cross-section means the simple model assuming a single current filament used so far is insufficient to calculate the trap parameters. Current is induced at a frequency of 30~kHz corresponding to in a skin depth of 0.4~mm in copper. This, combined with the radially dependent emf that scales as $r^2$ due to the increased magnetic flux enclosed in a larger area, confines the induced current to the outer edge of the ring. The exact distribution of current density and phase induced within the conductor is determined using a finite element simulation \cite{FEMM}, the magnitude of which is plotted in \fref{fig:fig2}(c) for $B_0=110$~G which reveals the strong current localisation, in good agreement with an independent lumped element calculation \cite{ryff70}. Integrating over the cross section gives a total induced current amplitude of 140~A, with phase $\delta_0$ corresponding to $\Omega=18$. The predicted power dissipated in the ring due to Ohmic heating is 4.3~W, giving an effective resistance of $440~\mu\Omega$ and $L=42.5$~nH, significantly larger than the dc values of 100~$\mu\Omega$ and 20~nH determined experimentally and corresponding to a uniform current distribution. The complete time-averaged trapping potential is then calculated using the theoretical current distribution to model the field from an array of $50\times20$ current filaments. \Fref{fig:fig2}(d) shows the trap potential for $B_0=110~$G and $B_\mathrm{b}=5$~G for the $\vert F=2, m_F=+2\rangle$ state of $^{87}$Rb including gravity to match experimental parameters presented below. The copper ring creates a trap at $r_\mathrm{trap}=5.2$~mm with radial and axial trap frequencies of 16~Hz and 60~Hz respectively. These frequencies are much slower than the 30~kHz ac frequency, validating the time-averaged assumption above. The figure also clearly shows both the region of avoided zeros either side of the trap minimum (yielding an equilibrium temperature of 4~$\mu$K), and the effects of gravitational sag that shifts the saddle point of the centre below the plane of the ring with a height of 740~$\mu$K. In the absence of Majorana losses, the total trap depth of the averaged magnetic potential is 2~mK.

\section{Experiment Setup}
\begin{figure}[b!]
\flushright
	\includegraphics{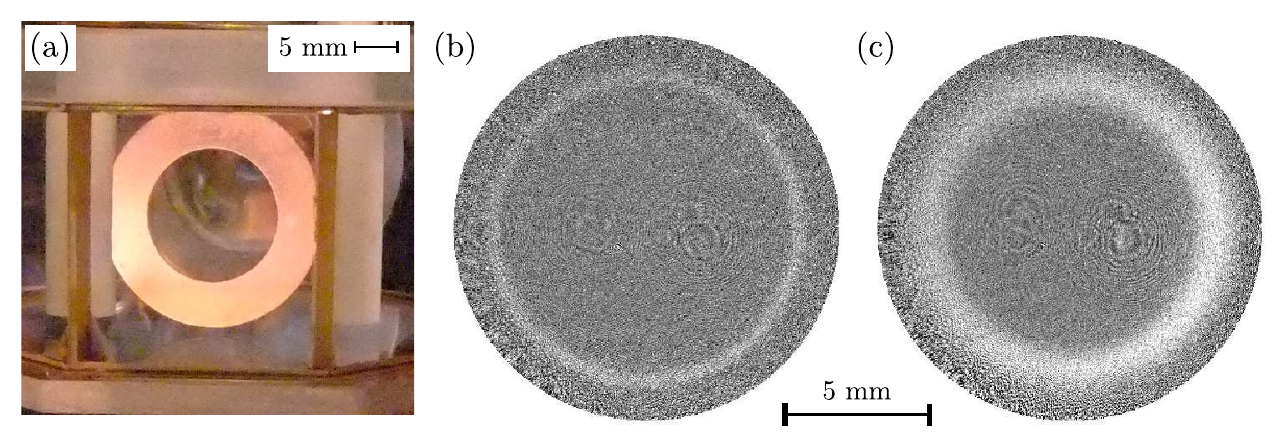}
	\caption{(a) Copper ring mounted in vacuo, supported by two macor bars, (b) 10-shot averaged absorption images of atoms released from the ring trap after 200~ms at $B_0=110$~G for $B_\mathrm{b}=4.6$~G with $N=0.5\times10^6$ and (c) $B_\mathrm{b}=9.2$~G with $N=3\times10^6$, demonstrating the change in the thickness of the traps as the zeros are pushed away from the trap minimum. The images are cropped to the internal diameter of the copper ring.}
	\label{fig:fig3}
\end{figure}

We experimentally characterise the time-averaged ring trap using a laser cooled cloud of $^{87}$Rb atoms, requiring the ring to be held in vacuum. The ring is mounted horizontally on a pair of Macor rods in a home-built octagonal glass vacuum cell, shown in \fref{fig:fig3}(a). The octagonal cell is constructed by gluing high quality BK-7 glass substrates to a glass-metal transition using Epotek 353ND to permit AR-coating on both sides of the glass. The ac drive field is provided by a pair of coils driven by a 600~W audio amplifier, using a series LCR resonance circuit tuned to 30.5~kHz to cancel the inductance of the drive coils and giving a maximum drive field of 110~G in the plane of the ring. The bias field is provided by shim coils surrounding the chamber, giving a maximum vertical bias of 9~G.

Atoms are cooled and trapped in a standard magneto-optical trap (MOT) that is axially centred 8~mm below the plane of the ring. Following a short 10~ms optical molasses to reduce the temperature to 20~$\mu$K, the atoms are optically pumped into the $\vert F=2, m_F=+2\rangle$ state and transferred into a 40~G/cm quadrupole trap. To load the atoms into the ring trap the quadrupole coils are used in conjunction with an additional bias coil aligned along the $\hat{x}$ axis to raise the atoms in the quadrupole trap to overlap the cloud with the ring trap radius $r_\mathrm{trap}\sim 5$~mm. Atoms are moved in 200~ms followed by a 10~ms hold time using a current ramp optimised to minimise heating, resulting in $1.1\times10^7$ atoms in the quadrupole trap with a temperature of 40~$\mu$K and radius $\sigma=0.6$~mm. The quadrupole trap is then turned off, and the ac drive coils and vertical bias field turned on. As a consequence of the coil geometry the quadrupole coils are strongly inductively coupled to the ac drive coils, and must be electrically isolated using an external relay that imposes a 0.5~ms delay between disabling the quadrupole and enabling the ac amplifier. The atoms then freely evolve in the ring trap for a variable time, before turning off the ring trap and performing absorption imaging of the atoms using a circularly polarised probe beam aligned along the $\hat{z}$-axis after a 3~ms time of flight. Due to the large and fluctuating Zeeman shift caused by the ac magnetic field amplitude it is not possible to image the atoms in the trap directly. 

One of the challenges of using a single-chamber vacuum system for the ac ring trap arises due to copper acting as an efficient getter material for rubidium atoms. This leads to a build up of rubidium atoms on the ring from the background vapour required for loading the MOT. At the peak ac drive amplitude there is over 4~W of power dissipated in the ring, leading to a heating rate of 2~K/s. The effect of this heating for long trap hold times or over accumulated experimental runs is to release rubidium from the copper surface, leading to a significant enhancement of the rubidium vapour pressure and consequently reducing the background limited lifetime in the trap and leading to large shot-to-shot atom number fluctuations. This issue was circumvented using low rubidium vapour pressures and long (8~s) MOT load times, with regular cleaning cycles performed by running the ac field at full power for up to an hour and waiting for the vapour pressure to recover. This problem might be overcome using UV light to perform light assisted atomic desorption (LIAD) \cite{klempt06} to prevent a build up of atoms on the copper ring, or coating the ring using an insulating material such as sapphire which acts as a less efficient getter of Rb.

\section{Results}

The theoretical analysis of the time-averaged ring potential presented in \sref{sec:theory} reveals the importance of the position of the instantaneous magnetic field zeros during each cycle of the ac field to avoid violating the adiabaticity requirement. To characterise this effect, data are taken for a range of ac field amplitudes $B_0$, which determine the initial trap depth, and bias fields $B_\mathrm{b}$ that control the adiabaticity of the ring potential.

\Fref{fig:fig3} shows absorption images of atoms after 200~ms evolution in the ring trap at $B_0 = 110$~G for $B_\mathrm{b} = 4.6$~G (b) and 9.2~G (c), with each image being the average of 10 repeats. These demonstrate the time-averaged potential creates a large radius, cylindrically symmetric waveguide for cold atoms. The effect of the vertical bias on the trap is clearly visible, with the width measured from the standard deviation of the radial distribution changing from a thin ring of $\sigma_r=0.19$~mm to a wide ring with $\sigma_r=0.51$~mm as the B-field zeros are pushed further from the trap centre, as illustrated in \fref{fig:fig2}(c). The magnetic field zeros act to expel hot atoms from the trap within a few cycles in a similar manner to the `circle of death' in a TOP trap, leading to an effective radial temperature of $7\pm0.5$~$\mu$K measured from time of flight expansion of atoms released from the ring. This is approximately twice the expected equilibrium temperature from \fref{fig:fig2}(d), likely due to the small fraction of a cycle the closest zero spends near the trap centre. However, due to the low density and hence low collision rate, there is no re-thermalisation within the ring potential and the atoms maintain an azimuthal velocity distribution corresponding to the 40~$\mu$K of the initial quadrupole trap making this technique ineffective for evaporation. Using a harmonic approximation for the bottom of the potential, the radial trap frequency can be estimated from the measured cloud size as $\omega_r=\sqrt{k_BT/m}/\sigma_r$, resulting in a trap frequency of $10\pm1$~Hz. As the bias is increased, the equilibrium temperature increases to $18\pm2~\mu$K for data in (b), above which the anharmonicity of the trap precludes accurate measurement of the radial temperature and trap frequency.

An additional consequence of the increasing bias field is a reduction in the axial trap frequency, leading to enhanced gravitational sag. This reduces the trap radius from $r_\mathrm{trap}=5.12$~mm in (b) to 4.84~mm in (c), which can be understood from the contour plot in \fref{fig:fig2}(d). A larger variation in radius is observed for smaller $B_0$ due to the reduction in the initial axial trap frequency. This behaviour shows good agreement with the finite element model discussed above for the trap parameters, enabling the atoms to be used as a probe of the magnetic field.

\begin{figure}[b!]
\flushright
	\includegraphics{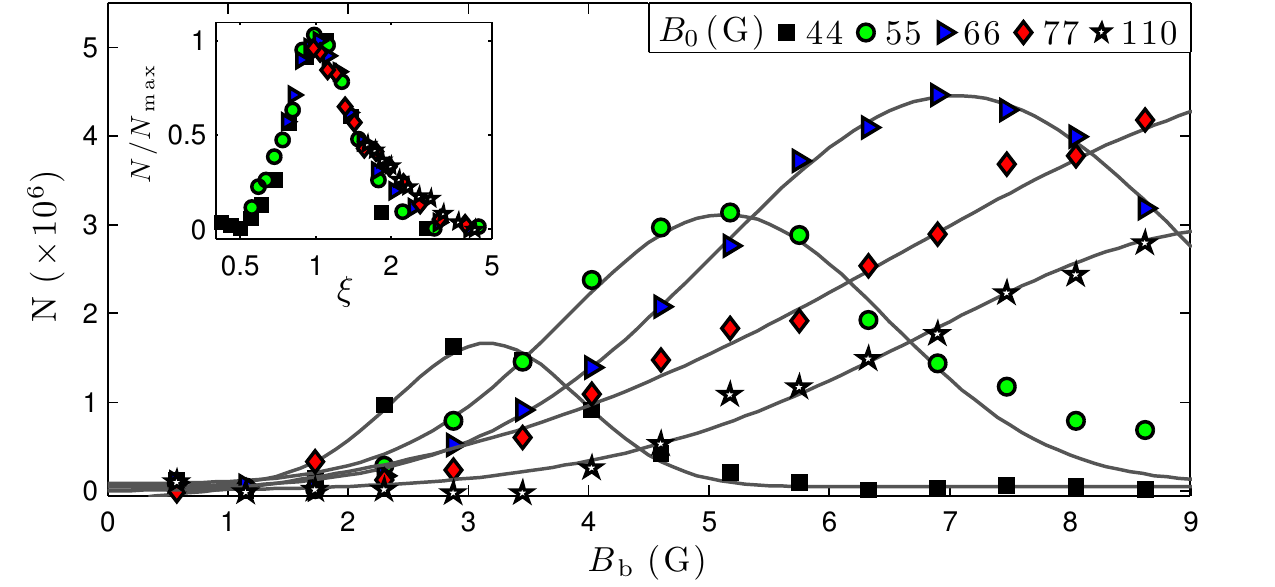}
	\caption{Ring trap characterisation of atom number in the ring trap after 200~ms as a function of applied bias field for different ac field amplitudes, $B_0$. Gaussian fits are shown as a guide to the eye, with errorbars equivalent to the markersize. Inset: Scaled data showing universal scaling of ring parameters with $\xi$ (see text).\label{fig:fig4}}	
\end{figure}

An important parameter in characterising the trap geometry is the loading efficiency from the quadrupole trap, and hence atom number within the trap. \Fref{fig:fig4} shows the atom number after 200~ms in the ring as a function of drive field and applied bias. This timescale is chosen to enable the atoms to spread round and completely fill the ring and to enable any untrapped atoms to fall out of the probe field of view. The peak atom number loaded into the ring at $B_0=66$~G is $4.5\times10^6$ corresponding to 43~\% of the initial number in the quadrupole trap, limited by the finite mode-matching between the quadrupole trap into the ring trap. Comparison of the required minimum bias field to the threshold value required to observe atoms in the ring trap shows good agreement with \eref{eq:Bmin} above, which predicts $B_\mathrm{b} > 3$~G for $B_0=55$~G. The variation of atom number with the bias field displays an approximately Gaussian dependence. The resulting trend can be understood qualitatively from the trade off between increasing the equilibrium temperature of the ring above that of the thermal cloud and loss of trap depth proportional to $B_\mathrm{b}$ which reduces the effective trapping volume of the ring. This leads to a universal scaling of the peak atom number within the trap in terms of parameter $\xi = (B_0-c)/(m\times B_\mathrm{b})$ as illustrated in the inset of \fref{fig:fig4}, where parameters $m=5.6$ and $c=26$~G are extracted from a straight line fit to the value $B_\mathrm{b}$ corresponding to the peak atom number as a function of $B_0$. The coefficients $m$ and $c$ give the scaling of bias field to drive field and the minimum ac amplitude required to obtain a ring trap respectively. Their exact values depend strongly on the temperature of the initial atomic sample, as if pre-cooled to a few micro-Kelvin the peak atom number should be achieved at low bias field without significant reduction in the trapping volume. This will also modify the shape of the profile for $\xi<1$ as a larger bias can be applied before the trap depth becomes comparable to the atomic kinetic energy.

\begin{figure}[b!]
\flushright
	\includegraphics{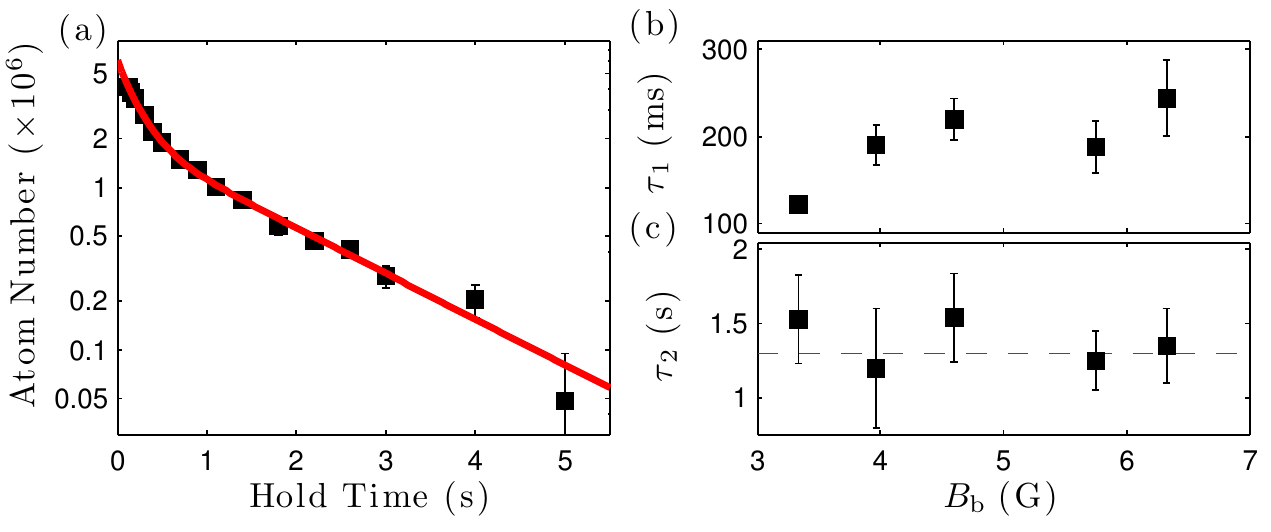}
	\caption{Ring Trap Lifetime at $B_0=55$~G. (a) Atom number at $B_z=4.6$~G vs hold time in the ring showing a two-component decay, with short timescale $\tau_1$ due to non-adiabatic losses and slow decay $\tau_2$ due to background collisions. (b) Lifetime $\tau_1$ vs $B_z$ plateaus around 220~ms as the trap is made adiabatic. (c) Background-limited lifetime $\tau_2$ is independent of bias, with quadrupole lifetime indicated by dashed line.\label{fig:fig5}}
\end{figure}

Analysis of the lifetime within the ring potential provides further evidence for this interpretation. \Fref{fig:fig5}(a) shows atom number as a function of hold time for $B_0=55$~G and a bias field of 4.6~G, showing a double exponential decay in the loss of atoms from the ring. Fitting the data allows extraction of the initially rapid fast decay time $\tau_1$, caused by non-adiabatic losses as hot atoms are evaporated out of the ring potential by the instantaneous zeros, and the longer timescale, $\tau_2$, associated with losses due to background collisions. \Fref{fig:fig5}(b) shows the change in $\tau_1$ as a function of bias field, which initially increases from 120~ms to 220~ms as the bias field increases to 5~G corresponding to an increase in the equilibrium temperature from 10 to 20~$\mu$K, in agreement with the increase in atom number for the data in \fref{fig:fig4}. There is no further gain for higher bias due to the relaxation of the trap which only slightly increases the equilibrium value. Importantly however, the background limited lifetime $\tau_2$ plotted in \fref{fig:fig5}(c) shows no dependence on the applied bias field, giving an average of $\tau_2 = 1.2\pm0.2$~s which matches the measured lifetime in the quadrupole trap. The biased ring trap therefore creates an adiabatic ring potential for cold atoms, permitting background limited lifetimes and hence long interaction times for atomic interferometry. Observation of the initial non-adiabatic loss is a direct consequence of the relatively hot thermal distribution loaded from the quadrupole trap.

As well as considering how the atom number changes in the ring potential, it is also interesting to consider the evolution of the atomic distribution within the circular waveguide. For atoms loaded into a thin ring with a relatively weak bias field, as seen in \fref{fig:fig3}(b), the Gaussian spatial distribution determined by the initial quadrupole trap spreads ballistically around the ring, taking 120~ms to completely fill the ring potential, which is determined by the 40~$\mu$K azimuthal temperature. At long times the azimuthal density distribution has a variation of 10~\%, independent of time. This corresponds to a potential difference of approximately 4~$\mu$K across the ring, consistent with a smooth trap tilted at an angle of 4~mrad. One of the proposed advantages of using ac currents for creating magnetic traps is to overcome issues of corrugation due to the electron motion in the conductor, however a significantly colder atomic sample is required to probe the smoothness of the potential on shorter length scales.

\begin{figure}
\flushright
	\includegraphics{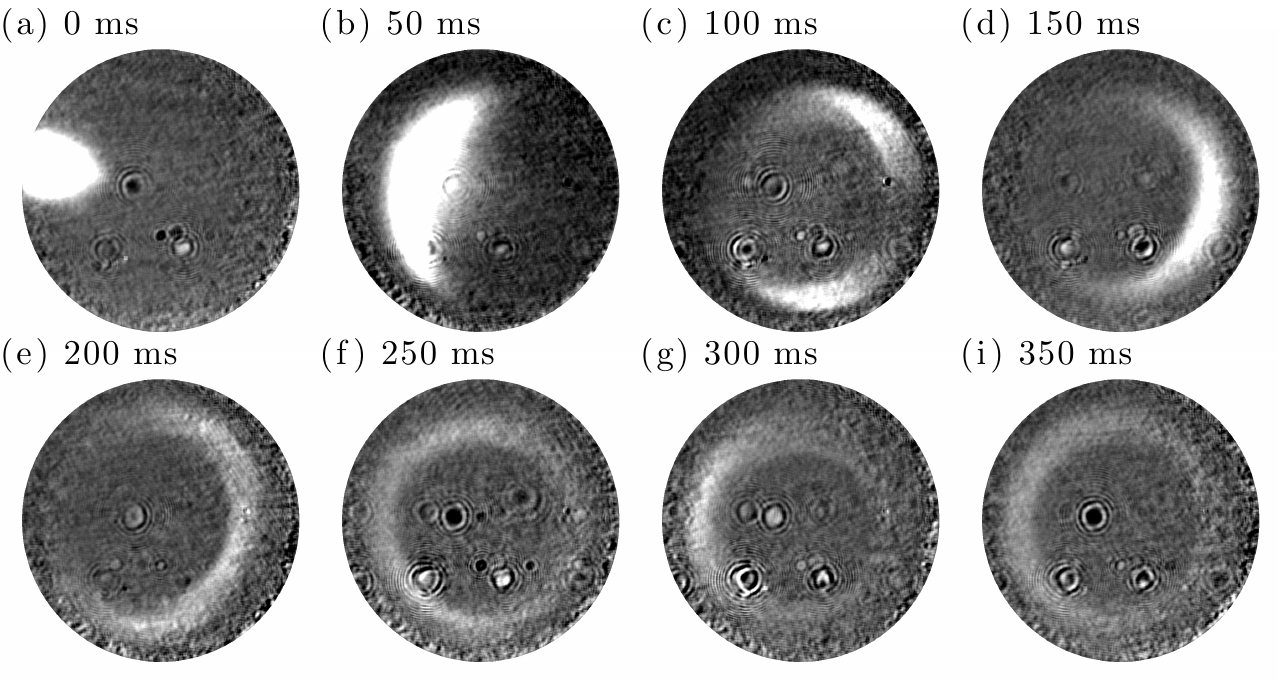}
	\caption{Evolution in the ring for $B_0=55$~G and $B_\mathrm{b}=4.6$~G. Atoms loaded on the edge of the ring trap can be split into two counter-propagating wave-packets using the curvature of the toroidal potential that spread out to fill the ring. The small circular fringes are due to imperfections in the imaging system.\label{fig:fig6}}
\end{figure}

For the wide ring geometry, with increased values of $B_\mathrm{b}$, the evolution in the ring is strongly dependent upon the initial loading position of the quadrupole trap and it is possible to induce radial or centre of mass oscillations for atoms in the ring trap due to the shallow potential. Another feature of the wide ring potential that can be exploited is the curvature of the central region seen from \fref{fig:fig2}(a), which can be used to act as a beam splitter for the atomic cloud. \Fref{fig:fig6} shows the evolution for atoms loaded on the outer edge of the trap for $B_0=55$~G and $B_\mathrm{b}=4.6~$G. Each image is the average of 10 repeats, which shows the cloud being accelerated into the ring centre due to the initial radial displacement and being split into two separate counter propagating clouds which overlap at 150~ms and then refocus at the initial loading position around 300~ms. Videos of the ring evolution for this data and for a weak bias are available on the group webpage \cite{ringvideos}. For interferometric applications the initial splitting can be achieved using an optical Bragg grating \cite{martin88} to reduce the centre of mass radial oscillations associated with this method, suppressing losses due to atoms exploring the non-adiabatic regions of the time-averaged potential.


\section{Outlook and Conclusion}
Characterisation of the trap properties as a function of the ac drive field and applied bias show the importance of meeting the requirements for adiabaticity by removing the instantaneous B-field zeros from the minimum of the trap. This leads to increased loading efficiency and a vacuum-limited lifetime at long times for atoms in the ring trap. Our current setup is limited by the relatively high initial temperature of the atomic sample and the 1~s background lifetime corresponding to a pressure of $10^{-9}$ Torr. However, a reduced pressure and using a colder sample or quantum degenerate gas would permit operation at a lower bias field. This makes it possible to create a tight harmonic radial confinement with radial trap frequencies of approximately 100~Hz to define a one-dimensional waveguide suitable for atom interferometry offering a large integration time. Our current ring has an area of $A=80$~mm$^2$, leading to a rotation sensitivity of $\delta\Omega = \hbar/(8mA\sqrt{N})=4$~nrad/s for a BEC of $N=10^5$ atoms in a single revolution. Extension to exploit the ring trap in this regime is the subject of further work.

We have presented the first demonstration of an inductively coupled ring trap for cold atoms which provides a viable technique for generating macroscopic toroidal waveguides for cold atoms. The main advantage over alternative approaches using current-carrying wires is that the trap potential is determined by the geometry of the conducting ring which can be machined to high tolerance, making it easy to define an axially symmetric trap without any end effects or distortions from the external coils. 

\ack
This work has been supported by the EPSRC under grant EP/G026068/1. P.F.G. received support from the RSE/Scottish Government Marie Curie Personal Research fellowship program

\section*{References}

\providecommand{\newblock}{}


\begin{thebibliography}{10}
\expandafter\ifx\csname url\endcsname\relax
  \def\url#1{{\tt #1}}\fi
\expandafter\ifx\csname urlprefix\endcsname\relax\def\urlprefix{URL }\fi
\providecommand{\eprint}[2][]{\url{#2}}

\bibitem{seaman07}
Seaman B~T, Kr\"amer M, Anderson D~Z and Holland M~J 2007 {\em Phys. Rev. A\/}
  {\bf 75} 023615

\bibitem{ramanathan11}
Ramanathan A, Wright K~C, Muniz S~R, Zelan M, Hill W~T, Lobb C~J, Helmerson K,
  Phillips W~D and Campbell G~K 2011 {\em Phys. Rev. Lett.\/} {\bf 106} 130401

\bibitem{moulder11}
{Moulder} S, {Beattie} S, {Smith} R~P, {Tammuz} N and {Hadzibabic} Z 2011
  \textit{Preprint} {cond-mat/1112.0334}

\bibitem{garay00}
Garay L~J, Anglin J~R, Cirac J~I and Zoller P 2000 {\em Phys. Rev. Lett.\/}
  {\bf 85} 4643

\bibitem{cronin09}
Cronin A~D, Schmiedmayer J and Pritchard D~E 2009 {\em Rev. Mod. Phys.\/} {\bf
  81} 1051--1129

\bibitem{gustavson97}
Gustavson T~L, Bouyer P and Kasevich M~A 1997 {\em Phys. Rev. Lett.\/} {\bf 78}
  2046

\bibitem{lenef97}
Lenef A, Hammond T~D, Smith E~T, Chapman M~S, Rubenstein R~A and Pritchard D~E
  1997 {\em Phys. Rev. Lett.\/} {\bf 78} 760

\bibitem{wu07}
Wu S, Su E and Prentiss M 2007 {\em Phys. Rev. Lett.\/} {\bf 99} 173201

\bibitem{canuel06}
Canuel B, Leduc F, Holleville D, Gauguet A, Fils J, Virdis A, Clairon A,
  Dimarcq N, Bord\'e C~J, Landragin A and Bouyer P 2006 {\em Phys. Rev.
  Lett.\/} {\bf 97} 010402

\bibitem{su10}
Su E~J, Wu S and Prentiss M~G 2010 {\em Phys. Rev. A\/} {\bf 81} 043631

\bibitem{gupta02}
Gupta S, Dieckmann K, Hadzibabic Z and Pritchard D~E 2002 {\em Phys. Rev.
  Lett.\/} {\bf 89} 140401

\bibitem{wang05}
Wang Y~J, Anderson D~Z, Bright V~M, Cornell E~A, Diot Q, Kishimoto T, Prentiss
  M, Saravanan R~A, Segal S~R and Wu S 2005 {\em Phys. Rev. Lett.\/} {\bf 94}
  090405

\bibitem{deissler08}
Deissler B, Hughes K~J, Burke J~H~T and Sackett C~A 2008 {\em Phys. Rev. A\/}
  {\bf 77} 031604

\bibitem{lan12}
Lan S~Y, Kuan P~C, Estey B, Haslinger P and M\"uller H 2012 {\em Phys. Rev.
  Lett.\/} {\bf 108} 090402

\bibitem{sagnac13}
Sagnac G 1913 {\em C. R. Hebd. Seances Acad. Sci.\/} {\bf 157} 708

\bibitem{arnold06}
Arnold A~S, Garvie C~S and Riis E 2006 {\em Phys. Rev. A\/} {\bf 73} 041606

\bibitem{sauer01}
Sauer J~A, Barrett M~D and Chapman M~S 2001 {\em Phys. Rev. Lett.\/} {\bf 87}
  270401

\bibitem{wu04}
Wu S, Rooijakkers W, Striehl P and Prentiss M 2004 {\em Phys. Rev. A\/} {\bf
  70} 013409

\bibitem{arnold04}
Arnold A~S 2004 {\em J. Phys. B\/} {\bf 37} L29

\bibitem{gupta05}
Gupta S, Murch K~W, Moore K~L, Purdy T~P and Stamper-Kurn D~M 2005 {\em Phys.
  Rev. Lett.\/} {\bf 95} 143201

\bibitem{zobay01}
Zobay O and Garraway B~M 2001 {\em Phys. Rev. Lett.\/} {\bf 86} 1195

\bibitem{lesanovsky06}
Lesanovsky I, Schumm T, Hofferberth S, Andersson L~M, Kr\"uger P and
  Schmiedmayer J 2006 {\em Phys. Rev. A\/} {\bf 73} 033619

\bibitem{vangeleyn12}
Vangeleyn M, Garraway B~M, Perrin H and Arnold A~S {In Preparation.}

\bibitem{lesanovsky07}
Lesanovsky I and von Klitzing W 2007 {\em Phys. Rev. Lett.\/} {\bf 99} 083001

\bibitem{sherlock11}
Sherlock B~E, Gildemeister M, Owen E, Nugent E and Foot C~J 2011 {\em Phys.
  Rev. A\/} {\bf 83} 043408

\bibitem{ryu07}
Ryu C, Andersen M~F, Clad\'e P, Natarajan V, Helmerson K and Phillips W~D 2007
  {\em Phys. Rev. Lett.\/} {\bf 99} 260401

\bibitem{bruce11}
Bruce G, Mayoh J, Smirne G, Torralbo-Campo L and Cassettari D 2011 {\em Phys.
  Scr.\/} {\bf T143} 014008

\bibitem{henderson09}
Henderson K, Ryu C, MacCormick C and Boshier M~G 2009 {\em New J. Phys.\/} {\bf
  11} 043030

\bibitem{west12}
West A~D, Wade C~G, Weatherill K~J and Hughes I~G 2012 {\em Appl. Phys.
  Lett.\/} {\bf 101} 023115

\bibitem{griffin08}
Griffin P~F, Riis E and Arnold A~S 2008 {\em Phys. Rev. A\/} {\bf 77} 051402

\bibitem{crookston05}
Crookston M~B, Baker P~M and Robinson M~P 2005 {\em J. Phys. B\/} {\bf 38} 3289

\bibitem{petrich95}
Petrich W, Anderson M~H, Ensher J~R and Cornell E~A 1995 {\em Phys. Rev.
  Lett.\/} {\bf 74} 3352

\bibitem{kraft02}
Kraft S, G\"unther A, Ott H, Wharam D, Zimmermann C and Fort\'agh J 2002 {\em
  J. Phys. B\/} {\bf 35} L469

\bibitem{leanhardt03}
Leanhardt A~E, Shin Y, Chikkatur A~P, Kielpinski D, Ketterle W and Pritchard
  D~E 2003 {\em Phys. Rev. Lett.\/} {\bf 90} 100404

\bibitem{rabi54}
Rabi I~I, Ramsey N~F and Schwinger J 1954 {\em Rev. Mod. Phys.\/} {\bf 26} 167

\bibitem{FEMM}
Meeker D~C {Finite Element Method Magnetics} v4.2 (April 2012)
  \url{http://www.femm.info}

\bibitem{ryff70}
Ryff P, Biringer P and Burke P 1970 {\em IEEE Trans. Power App. Syst.\/} {\bf
  PAS-89} 228 ISSN 0018-9510

\bibitem{klempt06}
Klempt C, van Zoest T, Henninger T, Topic O, Rasel E, Ertmer W and Arlt J 2006
  {\em Phys. Rev. A\/} {\bf 73} 013410

\bibitem{ringvideos}
 {Videos of ring evolution available at
  \url{http://photonics.phys.strath.ac.uk/atom-optics/inductive-trap-ring-2/ring-trap-videos/}}

\bibitem{martin88}
Marting P~J, Oldaker B~G, Miklich A~H and Pritchard D~E 1988 {\em Phys. Rev.
  Lett.\/} {\bf 60} 515

\end{thebibliography}
\end{document}